\documentclass[journal]{IEEEtran}
\usepackage{amsmath}
\usepackage{graphicx}
\usepackage{caption2}
\usepackage{amsthm}
\usepackage{setspace}
\usepackage{mathtools}
\usepackage{epstopdf}
\usepackage{subfigure}
\usepackage{amsfonts}
\usepackage{amssymb}
\begin{document}
\title{An Upper Bound for the Capacity of Amplitude-Constrained Scalar AWGN Channel}
\author{{Borzoo Rassouli and Bruno Clerckx} %
\thanks{Borzoo Rassouli and Bruno Clerckx are with the Communication and Signal Processing group of Department of Electrical and Electronics,
Imperial College London, United Kingdom. emails: \{b.rassouli12; b.clerckx\}@imperial.ac.uk}
\thanks{Bruno Clerckx is also with the School of Electrical Engineering, Korea University, Korea.}
}
\maketitle
\begin{abstract}
This paper slightly improves the upper bound in Thangaraj et al. for the capacity of the amplitude-constrained scalar AWGN channel. This improvement makes the upper bound within 0.002 bits of the capacity for $\frac{E_b}{N_0}\leq 2.5$ dB.
\end{abstract}
\begin{IEEEkeywords}
Capacity, upper bound, amplitude constraint
\end{IEEEkeywords}

\section{Introduction}
The capacity of the point-to-point communication system subject to amplitude and variance (or equivalently, peak and average power) constraints was
investigated in \cite{Smith} for the scalar Gaussian channel where it was shown that the capacity-achieving distribution is
unique and has a probability mass function with a finite number of mass points. Consequently, the capacity and its achieving distribution can be evaluated numerically where the number, position and probabilities of the mass points are obtained by means of computer programs.

In \cite{McKellips}, an analytic upper bound is provided for the capacity which reduces the computational burden of numerical methods significantly. Recently, the bound in \cite{McKellips} was refined in \cite{Thangaraj}. In this paper, this bound is further refined by means of increasing the number of optimization parameters. In other words, we observe that using a test density whose tails decay as those of a Gaussian distribution with a variance slightly less than one can tighten the upper bound.

The paper is organized as follows. Section \ref{system} provides some preliminaries helpful for the remainder of the paper. The main result of this paper is given as a theorem in section \ref{results}. A comparison of the bounds is provided in section \ref{numerical} followed by section \ref{conclusion} which concludes the paper.
\section{Preliminaries}\label{system}
For a memoryless channel with input $X$, output $Y$, input Cumulative Distribution Function (CDF) $F_X(x)$ with support $\mathbb{S}$ and the channel density $f_{Y|X}(y|x)$, we have
\begin{align}
    C &= \sup_{F_X(x)}I(X;Y)\nonumber\\
    &=\sup_{F_X(x)} \int{D\left(f_{Y|X}(.|x)||f_Y(.)\right)dF_X(x)}\label{x2}\\
    &\leq \sup_{F_X(x)} \int{D\left(f_{Y|X}(.|x)||q_Y(.)\right)dF_X(x)}\label{x3}\\
    &\leq \sup_{x\in\mathbb{S}}D\left(f_{Y|X}(.|x)||q_Y(.)\right)\label{x4}
\end{align}
where in (\ref{x2}), $D(a||b)$ denotes the relative entropy between the densities $a$ and $b$. The inequality in (\ref{x3}) is a direct consequence of the non-negativity of relative entropy, i.e. $D\left(f_Y(.)||q_Y(.)\right)\geq 0$ in which $q_Y(y)$ is an arbitrary test density. Note that, the more similar $q_Y(y)$ is to $f_Y(y)$, the tighter becomes the upper bound in (\ref{x3}).

For the scalar AWGN channel, we have
\begin{equation}\label{x1}
    Y = X + N
\end{equation}
where $N\sim \mathcal{N}(0,1)$ is a Gaussian noise independent of the input. The amplitude-constrained capacity of this channel is
\begin{equation}
    C = \max_{F_X(x):|X|\leq A}I(X;Y)\label{x5}
\end{equation}
where $A$ denotes the amplitude constraint.

It was shown in \cite{Smith} that the capacity-achieving distribution $F^*_X(x)$ has a finite number of mass points in $[-A,A]$. McKellips proposed an analytic upper bound for $C$ based on bounding the entropy of $Y$ in \cite{McKellips}. In \cite{Thangaraj}, the upper bound for the capacity is further refined. The main idea is to find a simple test density $q_Y(y)$ that looks quite similar to the optimal output density $f^*_Y(y)$, which results from the optimal input $F^*_X(x)$, and plug it into (\ref{x3}) to get a tight upper bound. Since, as mentioned before, the more similar $q_Y(y)$ is to $f_Y(y)$, the tighter becomes the upper bound in (\ref{x3}).
\begin{figure}[t]
  \centering
  \includegraphics[width=9cm]{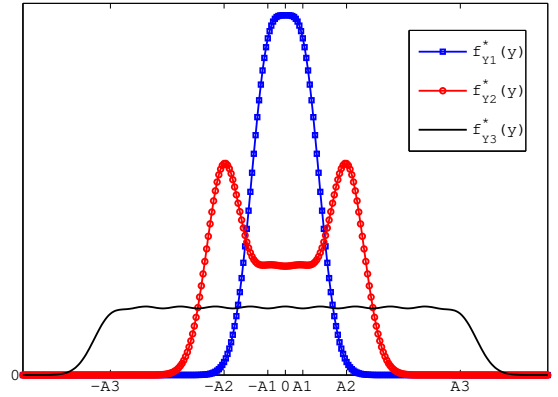}\\
  \caption{The optimal output density as $A$ increases.}\label{fig1}
\end{figure}

Figure \ref{fig1} shows the optimal output density $f^*_Y(y)$ for three values of the amplitude constraint ($A_1<A_2<A_3$). As it can be observed, it is intuitive to take a test density $q_Y(y)$ which is uniform on $[-A,A]$ and has Gaussian tails towards infinity\footnote{According to figure \ref{fig1}, this choice of test density is more acceptable in small or very large values of $A$.}.

The following functions are frequently used throughout the paper
\begin{align*}
    \psi(x) &= \frac{1}{\sqrt{2\pi}}e^{-\frac{x^2}{2}}\\
    Q(x) &= \int_{x}^{+\infty}\psi(t)dt\\
    g(u) &\triangleq u^2Q(u)-u\psi(u).
\end{align*}
For the capacity in (\ref{x5}), a trivial upper bound is the capacity with average power constraint, i.e. $\frac{1}{2}\log(1+P)$, in which $P=A^2$. Therefore, the bounds proposed in literature have the general form of
\begin{equation}\label{x5.5}
    C\leq \min\left\{\mathcal{T}(P),\frac{1}{2}\log(1+P)\right\}
\end{equation}
where in \cite{McKellips}, we have
\begin{equation}\label{x6}
    \mathcal{T}(P)=\log\left(1+\sqrt{\frac{2P}{\pi e}}\right)
\end{equation}
and in \cite{Thangaraj}, it was tightened further for $P\leq6.303$ dB as\footnote{This is the RHS of (17) in \cite{Thangaraj}.}
\begin{equation}\label{x7}
    \mathcal{T}(P)=\beta(P)\log\sqrt{\frac{2P}{\pi e}}+H(\beta(P))
\end{equation}
in which $\beta(P)=\frac{1}{2}-Q(2\sqrt{P})$ and $H(x)=-x\log(x)-(1-x)\log(1-x).$\footnote{Throughout the paper, the logarithms are in base $e$.}

In the following section, we further tighten $\mathcal{T}(P)$ for the whole SNR regime.
\section{Main results}\label{results}
\textbf{Theorem.} The capacity in (\ref{x5}) has the following upper bound
\begin{align}
    C\leq\min\left\{R(P)+W(P),\frac{1}{2}\log(1+P)\right\}\label{xx}
\end{align}
where
\begin{align}
    W(P) &= \frac{1}{2}\left(\log\sigma^{2}(P)+\frac{1}{\sigma^{2}(P)}-1\right)\left(\frac{1}{2}+Q(2\sqrt{P})\right)\nonumber\\
    &\ \ \ +\frac{g(2\sqrt{P})}{2\sigma^{2}(P)}\label{x8}
\end{align}
in which
\begin{equation}\label{x9}
    \sigma^{2}(P)= 1 + \frac{2g(2\sqrt{P})}{1+2Q(2\sqrt{P})},
\end{equation}
and
\begin{equation}\label{x10}
    R(P)=\left\{\begin{array}{cc} \log\left(1+\sqrt{\frac{2P}{\pi e}}\right) & P\geq 6.303\mbox{dB}\\ \beta(P)\log\sqrt{\frac{2P}{\pi e}}+H(\beta(P)) & \mbox{otherwise} \end{array}\right..
\end{equation}
Note that in the very small/large SNR regimes (i.e., $P\ll 0.1$ or $P\gg 0.5$), $\sigma^2(P)\approx 1$ and $g(2\sqrt{P})\approx 0$ which makes the bound boil down to (\ref{x6}) and (\ref{x7}).
\begin{proof}
Consider the following family of test densities
\begin{equation}
    q_Y(y)= \left\{\begin{array}{cc} \frac{\beta}{2A} & |y|\leq A\\ \frac{1-\beta}{\sqrt{2\pi\sigma^2}}e^{-\frac{(|y|-A)^2}{2\sigma^2}} & |y|> A \end{array}\right.
\end{equation}
where $\sigma^2$ and $\beta(\in[0,1])$ are parameters to be optimized. With this choice of test density, the relative entropy in (\ref{x4}) is evaluated as
\begin{align}
    & D\left(f_{Y|X}(.|x)||q_Y(.)\right)\nonumber\\
    &=\int_{-\infty}^{+\infty}\psi(y-x)\log\frac{\psi(y-x)}{q_Y(y)}dy\nonumber\\
    &=\underbrace{\log\frac{2A}{\beta\sqrt{2\pi e}}+\log\frac{\beta\sqrt{2\pi e}}{(1-\beta)2A}\left[Q(A-x)+Q(A+x)\right]}_{\mbox{RHS of (10) in }\cite{Thangaraj}}\nonumber\\
    &\ \ \ +\frac{1}{2}\left(\log\sigma^{2}+\frac{1}{\sigma^{2}}-1\right)\left[Q(A-x)+Q(A+x)\right]\nonumber\\
    &\ \ \ +\frac{1}{2\sigma^2}[g(A-x)+g(A+x)].\label{rhs}
\end{align}
We first find the maximum of (\ref{rhs}) over $x\in[-A,A]$ and then minimize this maximum value over the parameters $\beta$ and $\sigma^2$. In other words,
\begin{equation}\label{x11}
    C\leq \min_{\beta,\sigma^2}\max_{-A\leq x\leq A}D\left(f_{Y|X}(.|x)||q_Y(.)\right).
\end{equation}
As it can be observed, (\ref{rhs}) is an even function of $x$ which makes the region of interest as $x\in[0,A].$ Also, the optimization of the first two terms in (\ref{rhs}) was done in \cite{Thangaraj}. Therefore, we focus on the remaining terms.
\begin{figure*}[!t]
\normalsize
  \centering
  \includegraphics[width=15cm]{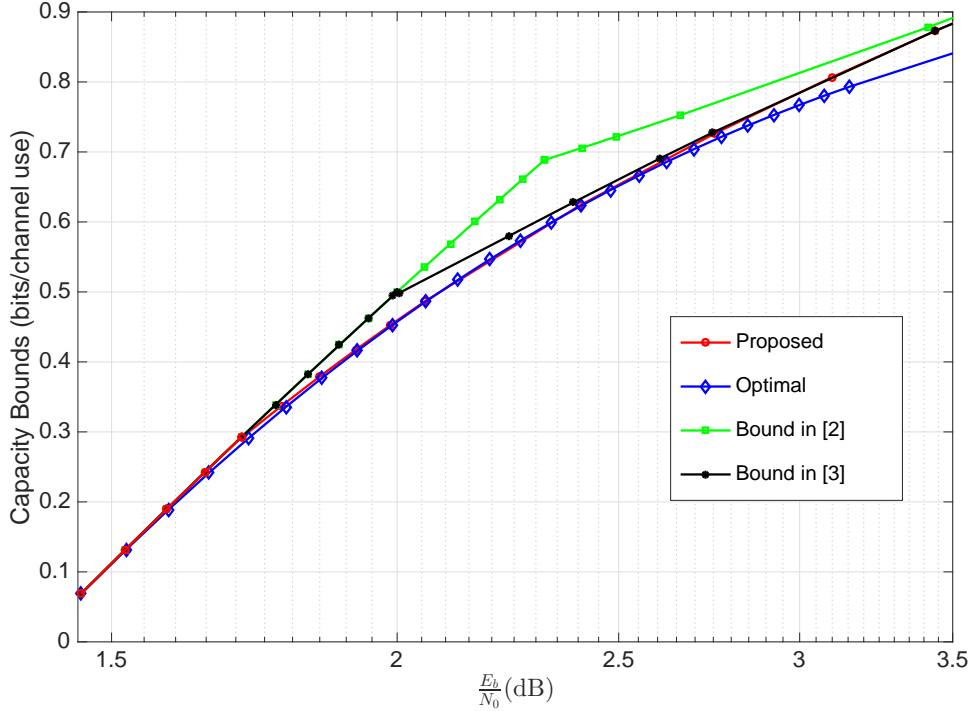}\\
  \caption{Comparison of the bounds.}\label{fig2}
\hrulefill
\vspace*{4pt}
\end{figure*}

\textbf{Lemma.} The following inequality holds for $\forall x\in[0,A]$
\begin{equation}\label{x12}
    g(A-x)+g(A+x)\leq g(2A).
\end{equation}
\begin{proof}
The proof is provided in Appendix.
\end{proof}
It can be easily verified that $Q(A-x)+Q(A+x)$ is an increasing function of $x\in[0,A]$ and $\log x+\frac{1}{x}-1\geq 0$ for $x> 0$. Therefore, using the lemma, we can write
\begin{align}
  &\frac{1}{2}\left(\log\sigma^{2}+\frac{1}{\sigma^{2}}-1\right)\left[Q(A-x)+Q(A+x)\right]\nonumber\\
    &+\frac{1}{2\sigma^2}[g(A-x)+g(A+x)]\nonumber\\
  &\leq \frac{1}{2}\left(\log\sigma^{2}+\frac{1}{\sigma^{2}}-1\right)\left(\frac{1}{2}+Q(2A)\right)+\frac{1}{2\sigma^2}g(2A).\label{x13}
\end{align}
The RHS of (\ref{x13}) is minimized by setting $\sigma^2$ as in (\ref{x9}) and the minimum is equal to $W(P)$ in (\ref{x8}). This completes the proof.
\end{proof}
Note that the lemma is the key part in allowing to add $\sigma^2$ to the optimization parameters, since if the trivial upper bound of zero is used instead of (\ref{x12}), the optimal value of $\sigma^2$ would be one (as used in \cite{McKellips} and \cite{Thangaraj}). 
\section{Numerical results}\label{numerical}
Figure \ref{fig2} compares the bounds in literature with the one proposed in this paper. Note that all the bounds are obtained by considering the minimum of two curves as in (\ref{x5.5}). We observe that the addition of $\sigma^2$ to the optimization problem results in a tighter bound. This small improvement of is mainly visible in the range $[1.5,2.5]$ dB (SNR per bit) as shown in the figure.
\section{Conclusion}\label{conclusion}
In this paper, the capacity of a scalar AWGN with amplitude-constrained input was considered and a further refinement of the upper bound in Thangaraj et al. was proposed. We observe that by optimizing over the variance of the test density, a tighter bound can be obtained.

Although the improvement is small, it can serve as a first step for looking at tighter bounds for the general vector AWGN channels which is of interest in optical communications. 
\appendices
\section{Proof of lemma}
Let
\begin{equation*}
    f_A(x) \triangleq g(A-x)+g(A+x)\ \ ,\ \ x\in[0,A].
\end{equation*}
For the function $g$, we can obtain the following properties
\begin{align}
    g(u)&\leq0\ \ ,\ \ u\geq 0 \label{ber1}\\
    g'(u) &\geq 0\ \ ,\ \ u\geq 1\label{ber2}.
\end{align}
(\ref{ber1}) is obtained as
\begin{align}
    g(u)&=u^2Q(u)-u\psi(u)\nonumber\\
    &< u\psi(u)-u\psi(u)\nonumber\\
    &=0\nonumber
\end{align}
where we have used the inequality $xQ(x)<\psi(x)$. (\ref{ber2}) is obtained as
\begin{align}
    g'(u) &= 2uQ(u)-\psi(u)\nonumber\\
    &>\frac{u^2-1}{u^2+1}\psi(u)\nonumber\\
    &\geq 0\ \ ,\ \ \mbox{for }u\geq 1\nonumber
\end{align}
where we have used the inequality $Q(x)>\frac{x\psi(x)}{1+x^2}$.

Therefore, for $A\geq 1$, we have
\begin{align}
    f_A(x)&<g(A+x)\label{ber3}\\
    &< g(2A)\label{ber4}
\end{align}
where (\ref{ber3}) and (\ref{ber4}) are due to (\ref{ber1}) and (\ref{ber2}), respectively.

For $A\leq 1$, we proceed as follows. The fourth derivative of $g$ is given by
\begin{align}
    \frac{d^4}{du^4}g(u)&=u(5-u^2)\psi(u)\nonumber
\end{align}
Hence, for $u\in [0,\sqrt{5})$, $\frac{d^4}{du^4}g(u)>0$ which indicates that $\frac{d^3}{du^3}g(u)$ is strictly increasing. This results in
\begin{equation}\label{x18}
    \frac{d^3}{dx^3}f_A(x)=\frac{d^3}{du^3}g(A+x)-\frac{d^3}{dx^3}g(A-x)\geq0
\end{equation}
for $A\leq 1$. (\ref{x18}) results in
\begin{align}
f''_A(x)&\geq f''_A(0)\nonumber\\
&= 2g''(A)\nonumber\\
&= 2[2Q(A)-A\psi(A)]\nonumber\\
&>\frac{2A(1-A^2)}{1+A^2}\psi(A)\label{x19}\\
&>0
\end{align}
where in (\ref{x19}), we have used the inequality $Q(x)>\frac{x\psi(x)}{1+x^2}$. Therefore, for $A\leq 1$, we have $f''_A(x)>0$ which results in $f'_A(x)>f'_A(0)=0$. Finally, having an increasing $f_A(x)$ confirms
\begin{equation*}
    f_A(x)<f_A(A)=g(2A).
\end{equation*}
This completes the proof.

\bibliography{REFERENCE}
\bibliographystyle{IEEEtran}

\end{document}